\documentstyle[emulateapj,psfig]{article}

  \def\itm#1 {\vskip10pt \noindent \square\ {\bf #1} }
  \def\square {\hbox{\vrule width5pt height5pt}}

  \def\deg      {{\ifmmode^\circ\else$^\circ$\fi} } 
  \def\arcm    {{\ifmmode {'\ }\else$'     $\fi} } 
  \def\arcs    {{\ifmmode{''\ }\else$''    $\fi} } 

\def\gs{\mathrel{\raise0.35ex\hbox{$\scriptstyle >$}\kern-0.6em
\lower0.40ex\hbox{{$\scriptstyle \sim$}}}}
\def\ls{\mathrel{\raise0.35ex\hbox{$\scriptstyle <$}\kern-0.6em
\lower0.40ex\hbox{{$\scriptstyle \sim$}}}}

\makeatletter

\newenvironment{inlinefigure}{%
\def\@captype{figure}%
\noindent\begin{minipage}{0.999\linewidth}\begin{center}}
{\end{center}\end{minipage}\smallskip}
\makeatother

\lefthead{Chapman et al.}
\righthead{MERLIN/VLA and HST imaging of Submm Galaxies}
\submitted{Astrophysical Journal, 614, 671-678, 2004 October 20}

\begin{document}
\title{Evidence for extended, obscured starbursts in submm galaxies}
\author{S.\,C.\ Chapman,$\!$\altaffilmark{1}
Ian Smail,\altaffilmark{2}
R.\ Windhorst,$\!$\altaffilmark{3}
T.\ Muxlow$\!$\altaffilmark{4}
\& R.\,J.\ Ivison\altaffilmark{5}
}

\altaffiltext{1}{California Institute of Technology, MS 320-47, Pasadena, CA, 91125}
\altaffiltext{2}{Institute for Computational Cosmology, University of Durham, South Road, Durham DH1 3LE, UK}
\altaffiltext{3}{Arizona State University, Dept.\ of Physics and Astronomy,
Tempe, AZ, 85287-1504}
\altaffiltext{4}{University of Manchester, MERLIN/VLBI National Facility, Jodrell Bank 
Observatory, Cheshire, SK11 9DL, UK}
\altaffiltext{5}{Astronomy Technology Centre, Royal Observatory, Blackford Hill, Edinburgh EH9 3HJ, UK}

\slugcomment{Received 2004 January 7; accepted 2004 June 29}

\begin{abstract}
We compare high-resolution optical and radio imaging of 12 luminous
submillimeter (submm) galaxies at a median $z= 2.2\pm 0.2$ observed
with {\it Hubble Space Telescope} ({\it HST}) and the MERLIN and VLA
radio interferometers at comparable spatial resolution, $\sim 0.3''$
($\sim 2$\,kpc).  The radio emission is used as a tracer of the likely
far-infrared morphology of these dusty, luminous galaxies. In $\sim
$30\% of the sample the radio emission appears unresolved at this
spatial scale, suggesting that the power source is compact and may
either be an obscured AGN or a compact nuclear starburst.  However, in
the majority of the galaxies, $\sim $70\% (8/12), we find that the
radio emission is resolved by MERLIN/VLA on scales of $\sim 1$\arcsec
($\sim$10\,kpc).  For these galaxies we also find that the radio
morphologies are often broadly similar to their restframe UV emission
traced by our {\it HST} imaging.  To assess whether the radio emission
may be extended on even larger scales, $\gg 1$\arcsec, resolved out by
the MERLIN+VLA synthesized images, we compare VLA B-array (5\arcsec\
beam) to VLA A-array (1.5\arcsec\ beam) fluxes for a sample of 50
$\mu$Jy radio sources, including 5 submm galaxies.  The submm galaxies
have comparable fluxes at these resolutions and we conclude that the
typical radio emitting region in these galaxies are unlikely to be much
larger than $\sim 1$\arcsec($\sim$10\,kpc). We discuss the probable
mechanisms for the extended emission in these galaxies and conclude
that their luminous radio and submm emission arises from a large,
spatially-extended starburst.  The median star formation rates for
these galaxies are $\sim 1700$\,M$_\odot$\,yr$^{-1}$ (M$>0.1$M$_\odot$)
occuring within regions with typical sizes of $\sim$40\,kpc$^2$, giving
a star formation density of 45\,M$_\odot$\,yr$^{-1}$\,kpc$^{-2}$.  Such
vigorous and extended starburst appear to be uniquely associated with
the submm population.  A more detailed comparison of the distribution
of UV and radio emission in these systems shows that the broad
similarities on large scales are not carried through to smaller scales,
where there is rarely a one-to-one correspondance between the
structures seen in the two wavebands.  We interpret these differences
as resulting from highly structured internal obscuration within the
submm galaxies, suggesting that their vigorous activity is producing
wind-blown channels through their obscuring dust clouds.  If correct
this underlines the difficulty of using UV morphologies to understand
structural properties of this population and also may explain the
surprising frequency of Ly$\alpha$ emission in the spectra of these
very dusty galaxies.
\end{abstract}


\section{Introduction}
\label{secintro}

Since their discovery, luminous submm galaxies (SMGs) have been
proposed as candidates for the progenitors of the most massive
spheroids in the local Universe (Smail, Ivison \& Blain 1997; Hughes et
al.\ 1998; Lilly et al.\ 1999; Blain et al.\ 2002).  The recent
measurement of the redshift distribution, space densities and
clustering of this population provides strong support for this proposed
relationship (Chapman et al.\ 2003a, 2004; Blain et al.\ 2004).  These
galaxies have large bolometric luminosities, $\sim$10$^{12}$--10$^{13}
L_\odot$, characteristic of ultraluminous infrared galaxies (ULIRGs,
Sanders \& Mirabel 1996).  If their intense restframe far-infrared
(far-IR) emission arise from dust-obscured star formation, then the
estimated rates are $\gs 10^3$\,M$_\odot$\,yr$^{-1}$, sufficient to
form the stellar population of a massive elliptical galaxy in only a
few dynamical times, given a sufficient gas reservoir.

Alternatively, a substantial fraction of the submm emission in these
galaxies could arise from an obscured AGN (e.g.\ Almaini et al.\ 1999).
It has proved difficult to distinguish whether AGN or starburst
activity powers the dust heating and associated far-IR radiation in
these luminous submm galaxies (Frayer et al.\ 1998; Alexander et al.\
2003; Chapman et al.\ 2003a). Optical and near-infrared spectroscopy or
X-ray observations have frequently been used to search for the
signatures of AGN, in both local ULIRGs and those at high-redshifts
(Sanders \& Mirabel 1996; Fabian et al.\ 2000; Ivison et al.\ 2000;
Barger et al.\ 2001; Frayer et al.\ 2003; Chapman et al.\ 2003a;
Alexander et al.\ 2003, 2004; Swinbank et al.\ 2004).  However, merely
identifying the presence of an AGN within a ULIRG does not immediately
mean that it must be the dominant source of far-IR radiation. Energetic
arguments must be used to estimate what fraction of a ULIRG's
luminosity arises from the AGN.

A much simpler test is available if the far-IR emission is resolved --
the geometry of the emission from an AGN means it is not a natural
source to heat dust over an extended region -- hence any extended
far-IR emission is very likely to arise from star formation.
Unfortunately, the coarse resolution of most far-IR and submm
instruments, e.g.\ $\sim 15''$ FWHM at 850\,$\mu$m with SCUBA on the
JCMT, means that the emission is rarely resolved except in the most
local galaxies (Le Floc'h et al.\ 2002). There have been recent claims
for the detection of submm emission on $\sim 100$\,kpc scales around
some powerful high-redshift AGN (Ivison et al.\ 2000; Stevens et al.\
2003, 2004), however, these are rare and extreme objects whose
characteristics may have little bearing on those of typical submm
galaxies. To disentangle the mechanisms responsible for the far-IR
emission in the population of submm galaxies at $z\sim 2$--3 (Chapman
et al.\ 2003a), will likely require sub-arcsecond resolution to map
emission on kpc scales -- well beyond the capabilities of current
far-IR/submm facilities.

One way to circumvent the limited spatial resolution of far-IR/submm
instruments is by exploiting the tight far-IR--radio correlation
observed for infrared galaxies (e.g.\ Helou et al.\ 1985; Condon 1992)
and the high angular resolution capabilites of long-baseline radio
inteferometers, such as the Multi-Element Microwave Linked
Interferometer (MERLIN) or the Very Large Array (VLA), to infer the
sub-arcsecond distribution of far-IR emission within submm galaxies.
One caveat of this approach is that the far-IR--radio correlation has
only been demonstrated locally on relatively large scales,
$\sim50$\,kpc (Yun, Reddy \& Condon 2001), and it the precise
correlation may break down on the smallest scales (M.\ Yun, in
preparation).  Nevertheless, if a significant extended component of the
continuum radio emission from the submm population is seen then this
would provide strong support for the far-IR emission being similarly
extended.

The deepest 1.4-GHz observations from the VLA can detect the
synchrotron emitting disks and nuclear starbursts (formed from the
coalescence of radio supernovae and their remnants) of a ULIRG such as
Arp\,220 out to redshifts of $z=3$--4.  Indeed, a large fraction of the
bright submm population at high redshifts are detected as $\mu$Jy radio
sources (Smail et al.\ 2000; Barger, Cowie \& Richards 2000; Chapman et
al.\ 2001; Ivison et al.\ 2002), as expected given their submm
luminosities and the local far-IR--radio correlation (Chapman et al.\
2004).  The highest spatial resolution available from the VLA at
1.4\,GHz is 1.5\arcsec, but by combining deep 1.4-GHz observations from
MERLIN and the VLA it is possible to produce datasets which combine
both high sensitivity and high spatial resolution, $\sim0.3$\arcsec\
scales, sufficient to map $\mu$Jy radio sources and test the extent of
the far-IR activity in these galaxies.

These sub-arcsecond radio maps, which indirectly trace the far-IR
morphologies, of the SMG population can also be compared and contrasted
with the restframe UV structures visible in {\it HST} imaging on
similar scales.  Such comparisons may help to constrain the extent of
obscuration in SMGs relative to other samples of high-redshift sources
selected in the restframe UV (e.g.\ Adelberger \& Steidel 2000).

In this paper, we present sensitive MERLIN/VLA radio and {\it HST}
restframe UV observations of SMGs at comparable, sub-arcsecond
resolution.  We discuss the sample and observations in \S2, describe
our main results in \S3 and discuss our conclusions in \S4.  We assume
a $\Lambda$-CDM cosmology with $\Omega_0=0.3$, $\Omega_\Lambda=0.7$ and
$H_0=65$\,km\,s$^{-1}$\,Mpc$^{-1}$, so that 1\,arcsec corresponds to
8.4\,kpc physical size at $z=2.2$.

%
%
\begin{figure*}[htb]
\centerline{
\psfig{figure=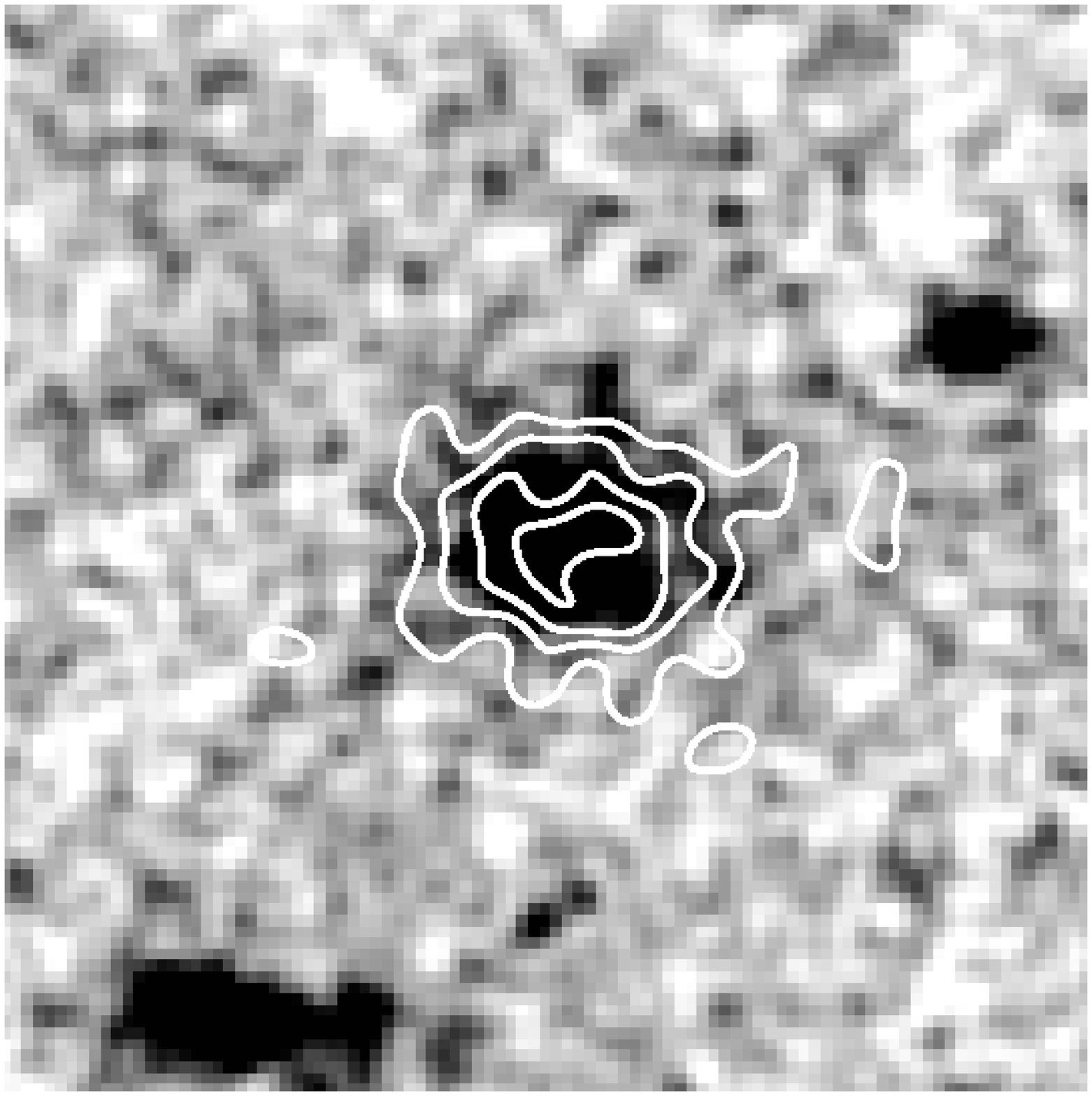,angle=0,width=1.6in,height=1.6in}
\psfig{figure=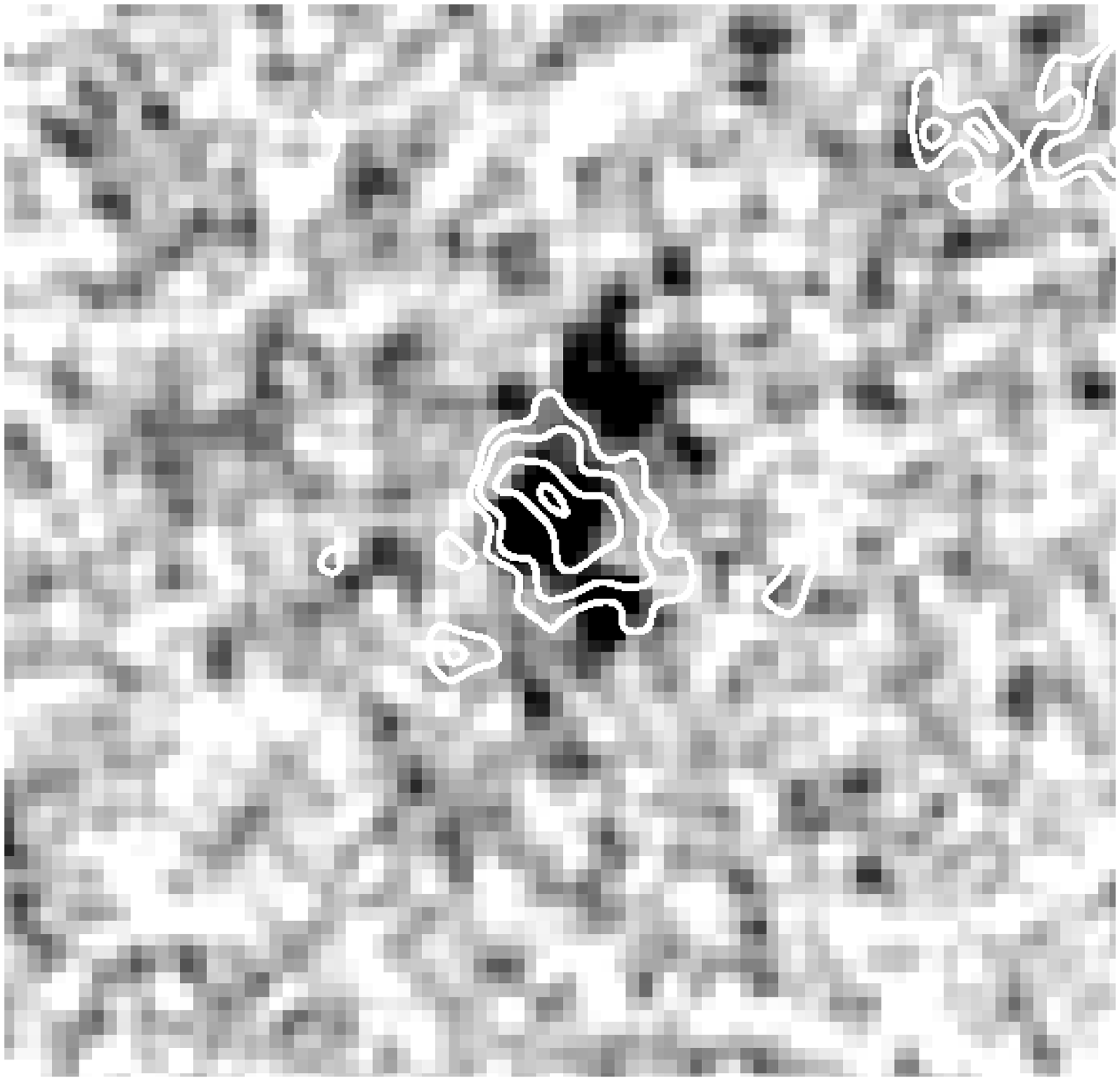,angle=0,width=1.6in,height=1.6in}
\psfig{figure=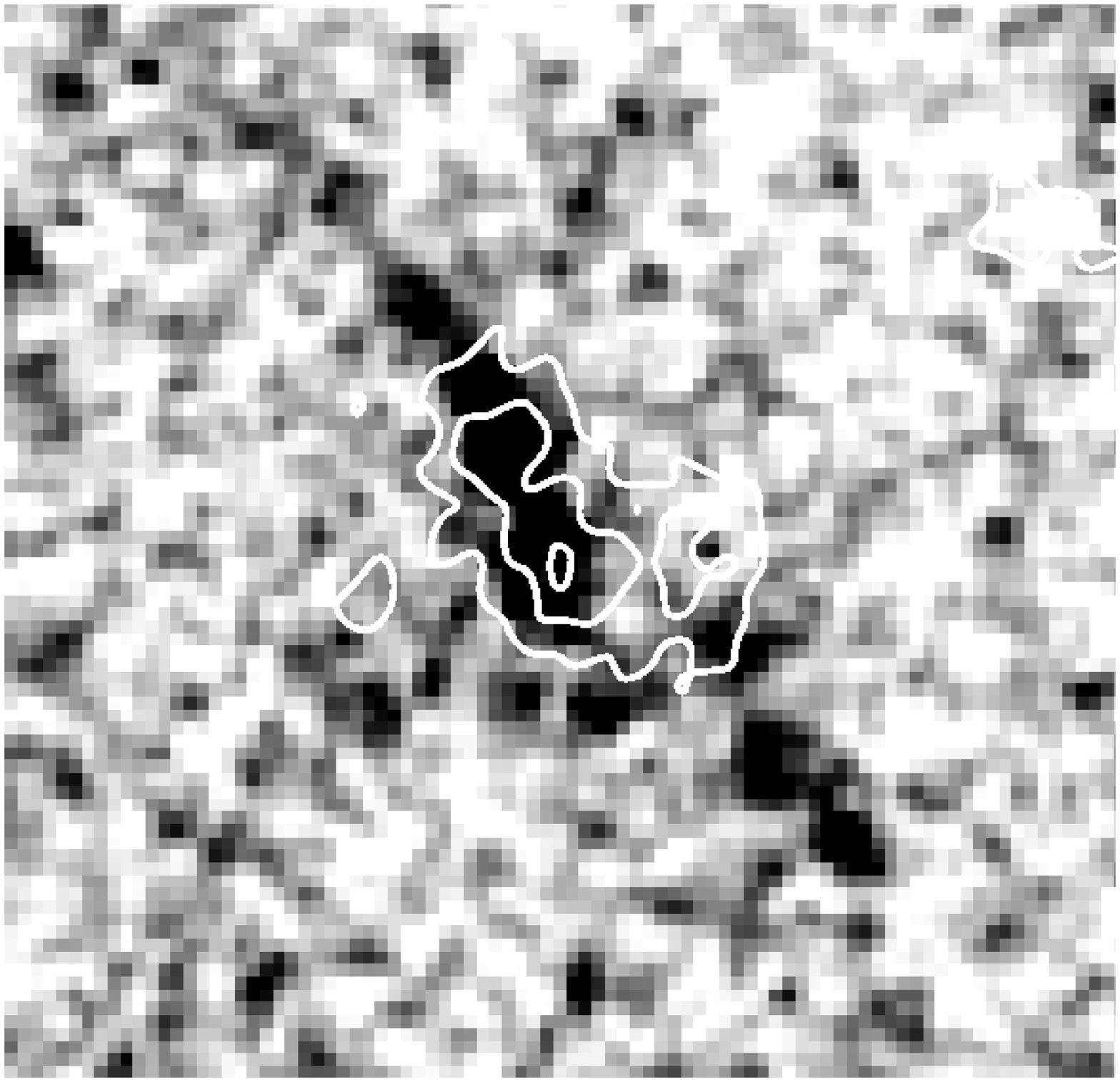,angle=0,width=1.6in,height=1.6in}
\psfig{figure=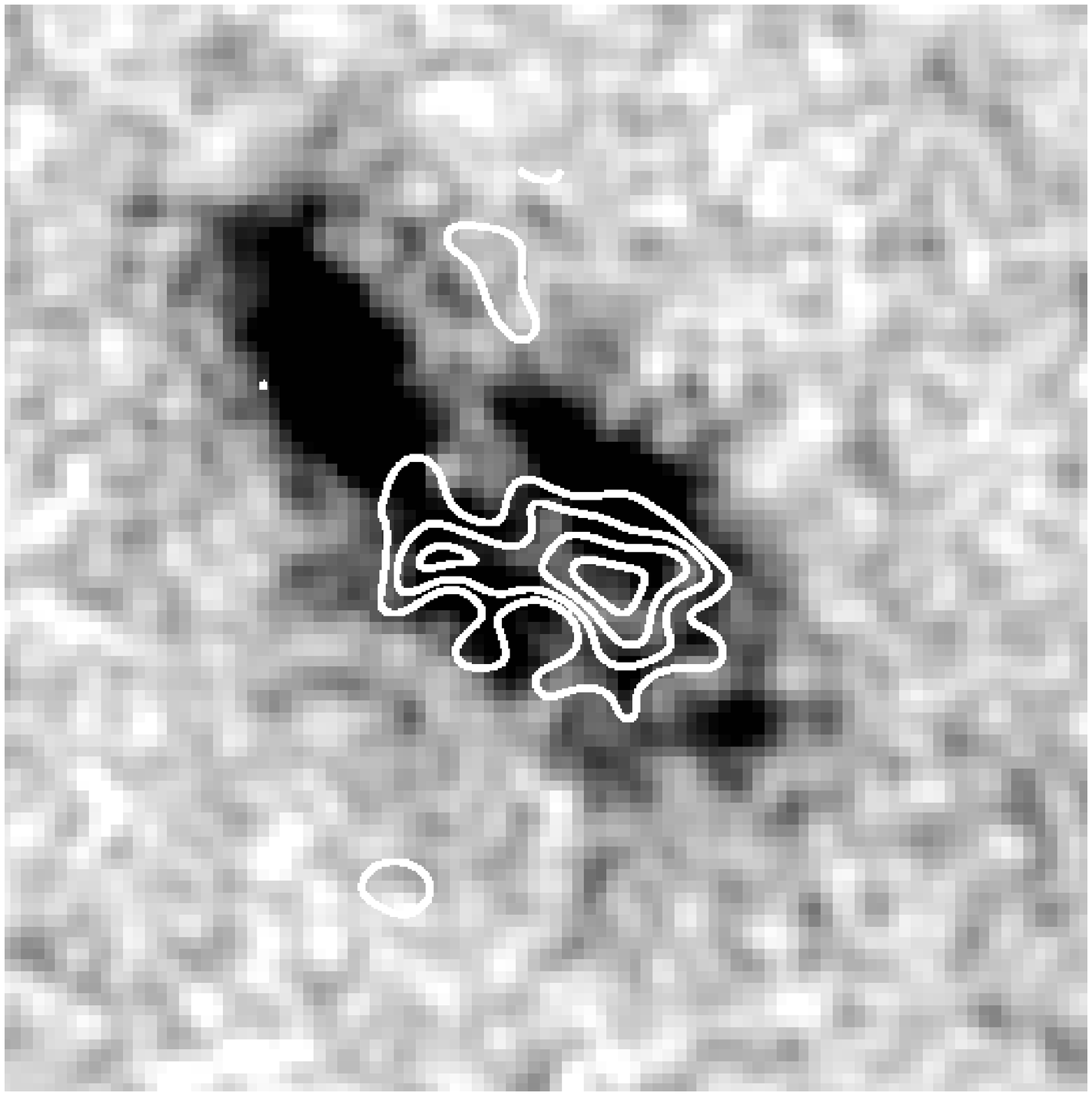,angle=0,width=1.6in,height=1.6in}
}
\centerline{
\psfig{figure=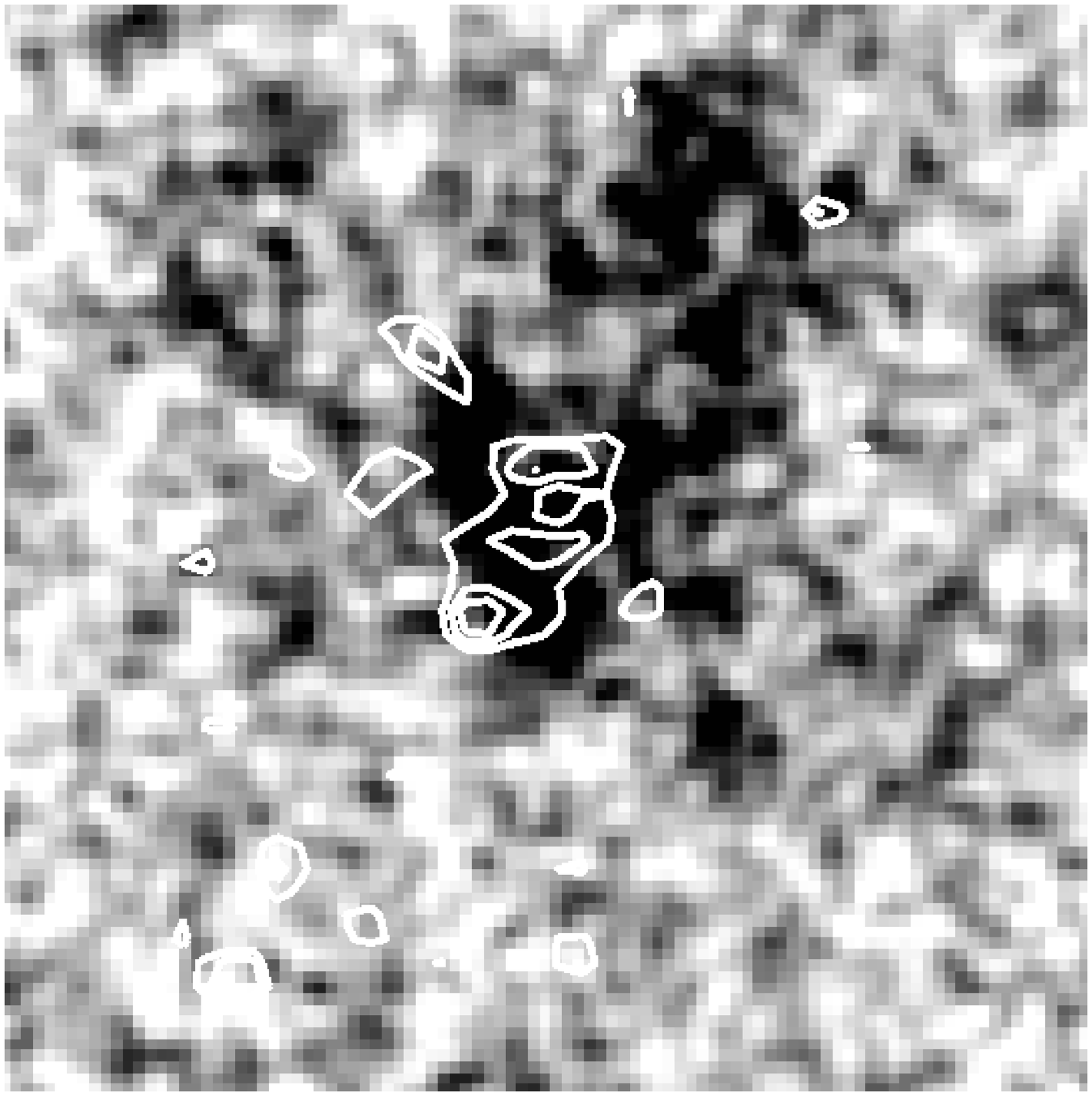,angle=0,width=1.6in,height=1.6in}
\psfig{figure=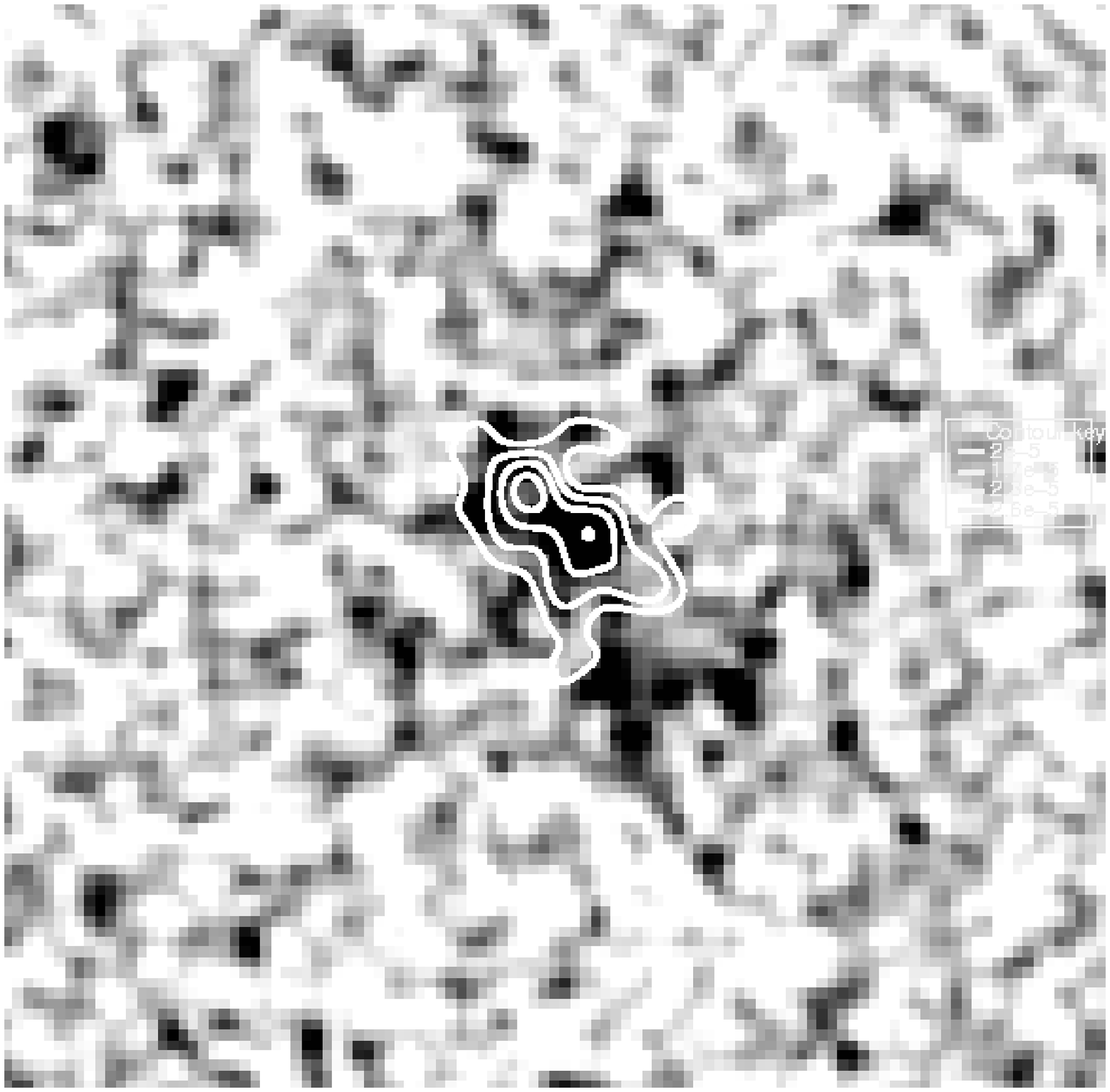,angle=0,width=1.6in,height=1.6in}
\psfig{figure=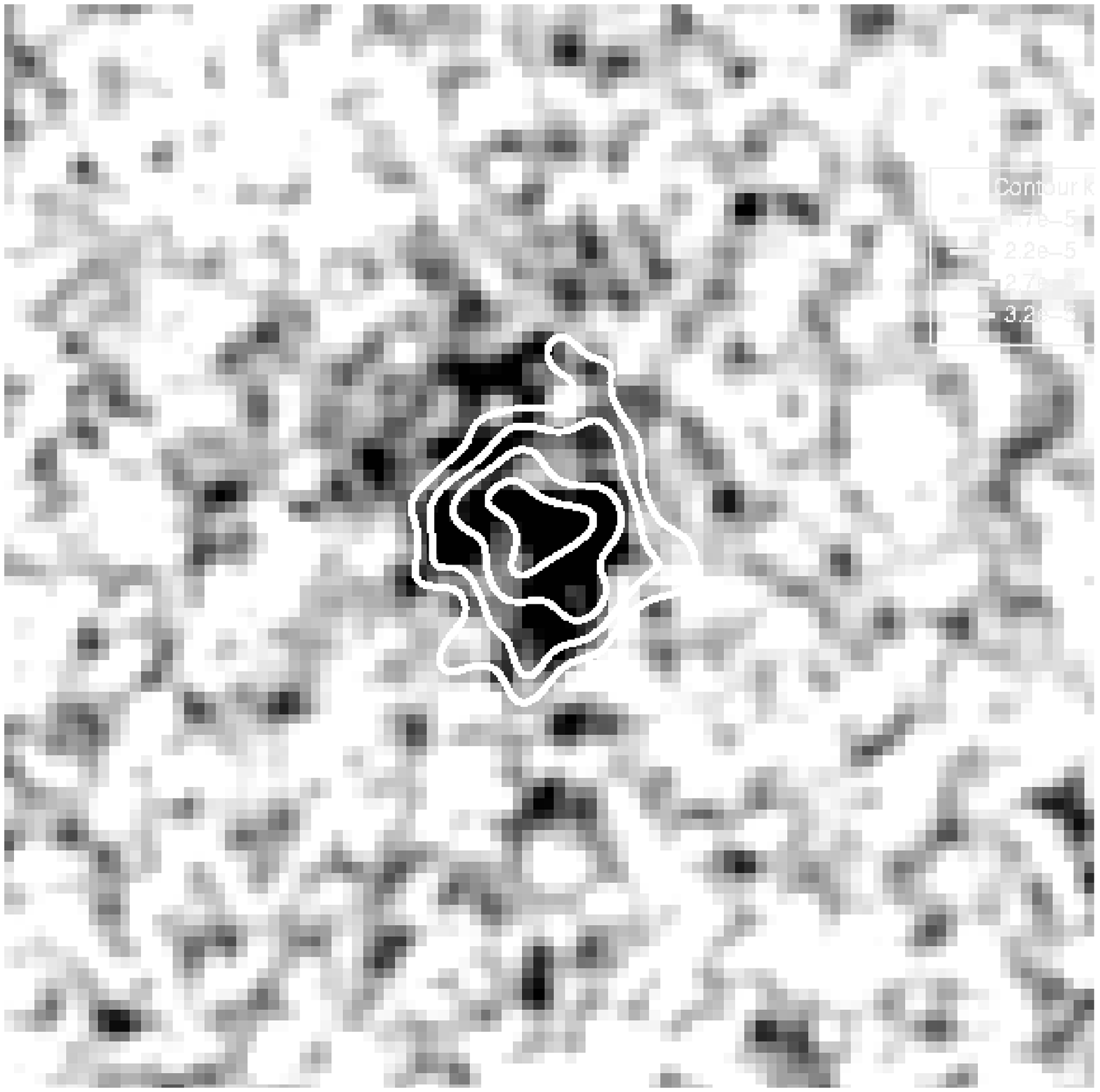,angle=0,width=1.6in,height=1.6in}
\psfig{figure=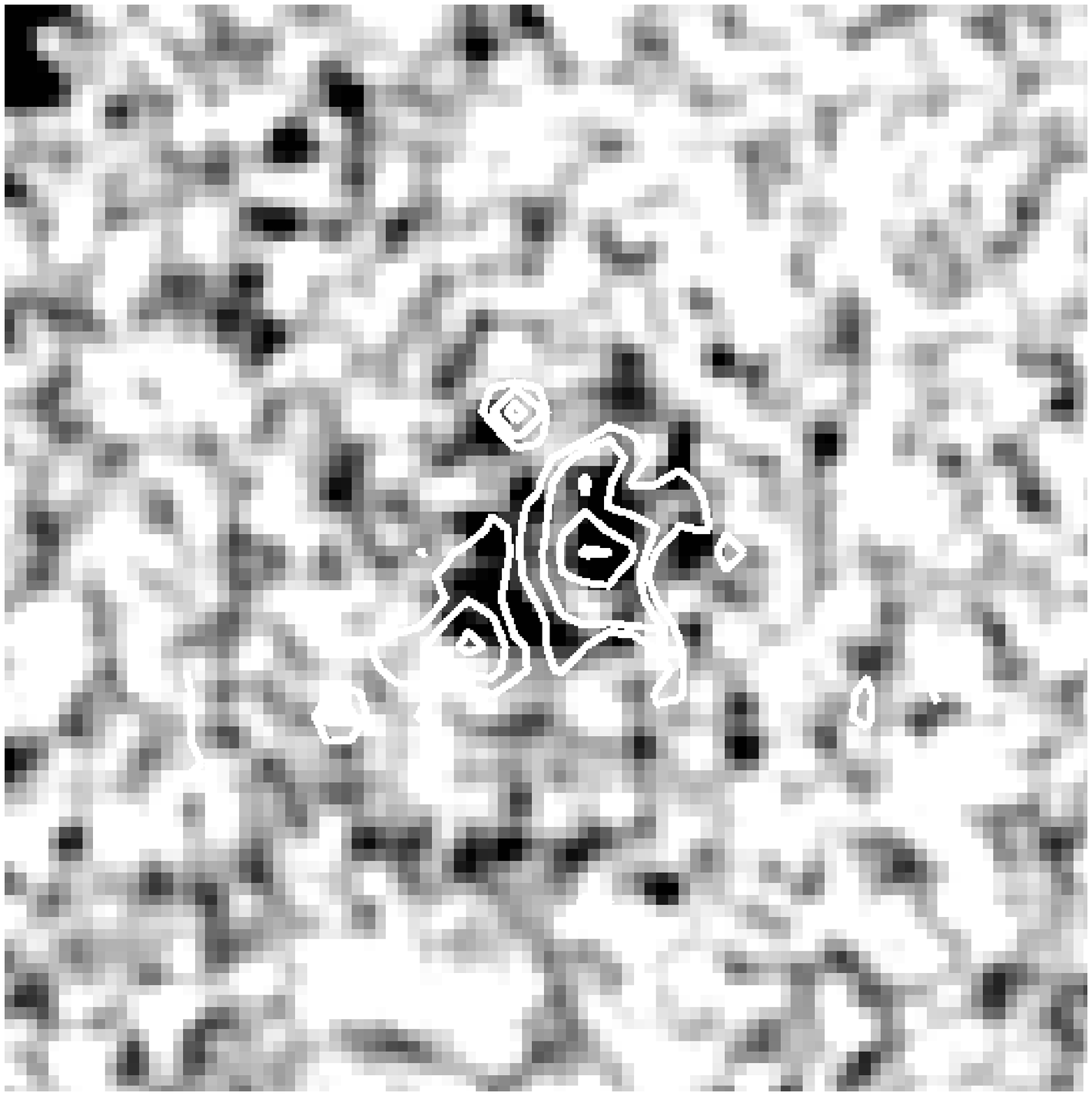,angle=0,width=1.6in,height=1.6in}
}
\centerline{
\psfig{figure=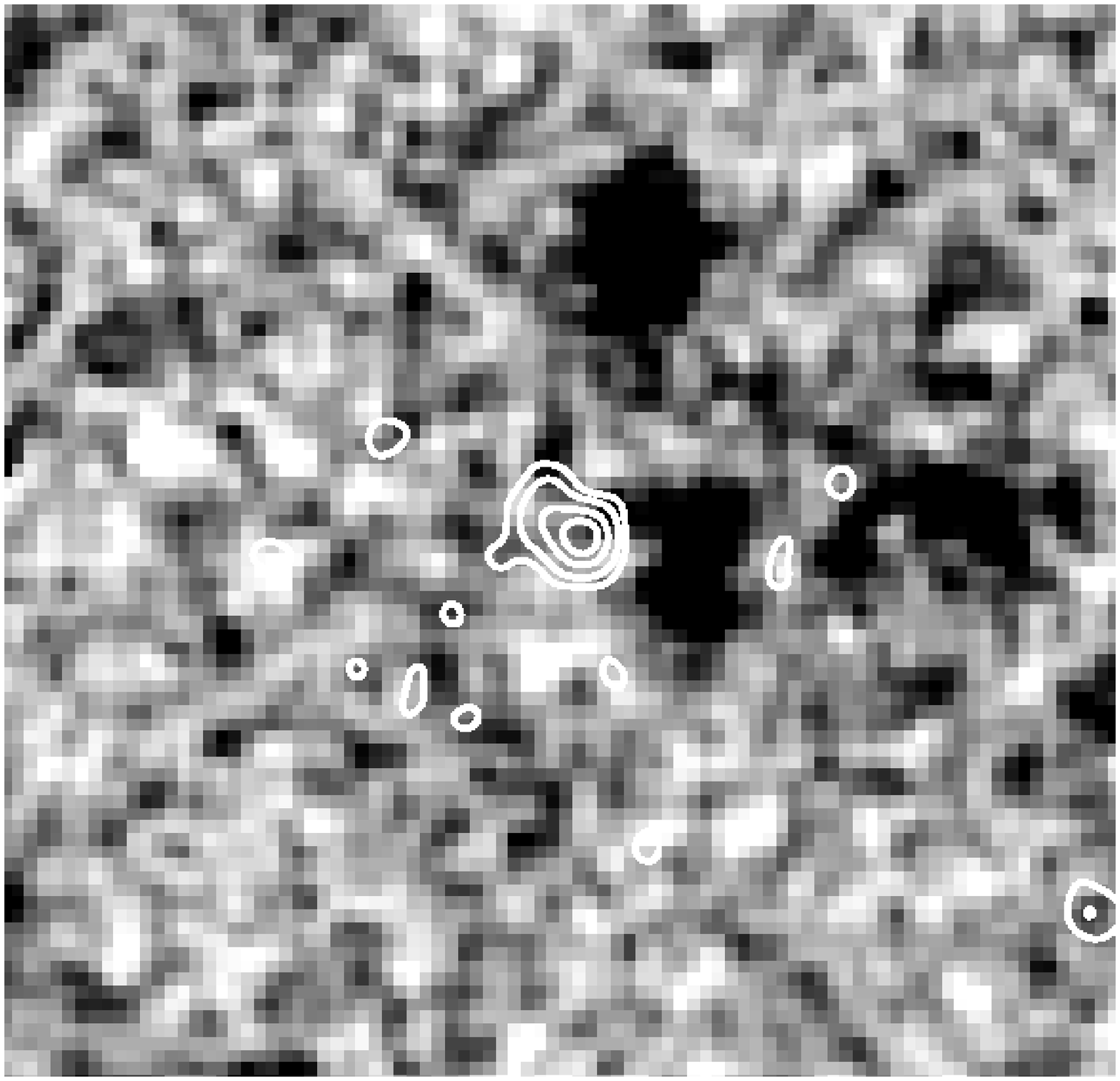,angle=0,width=1.6in,height=1.6in}
\psfig{figure=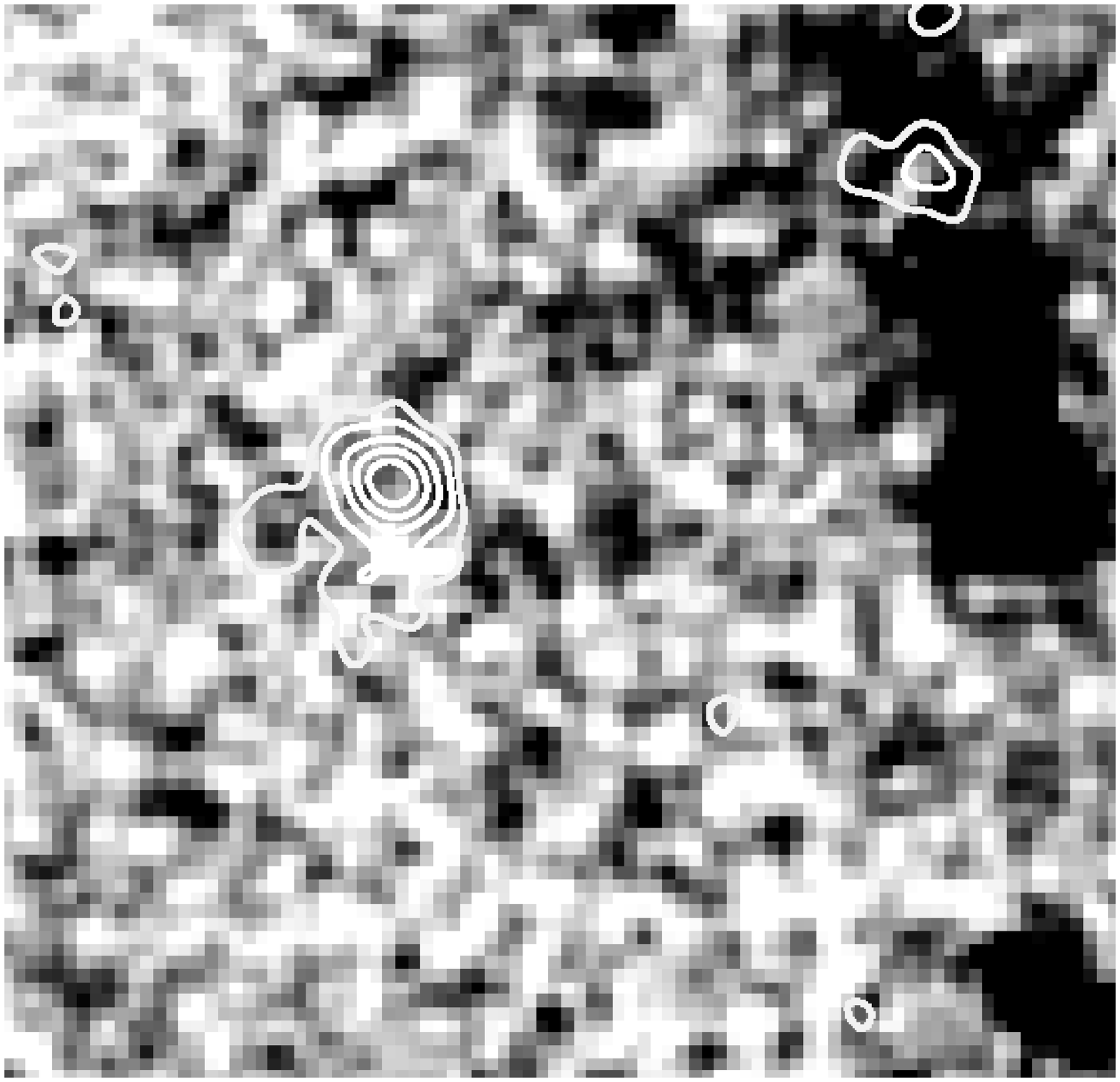,angle=0,width=1.6in,height=1.6in}
\psfig{figure=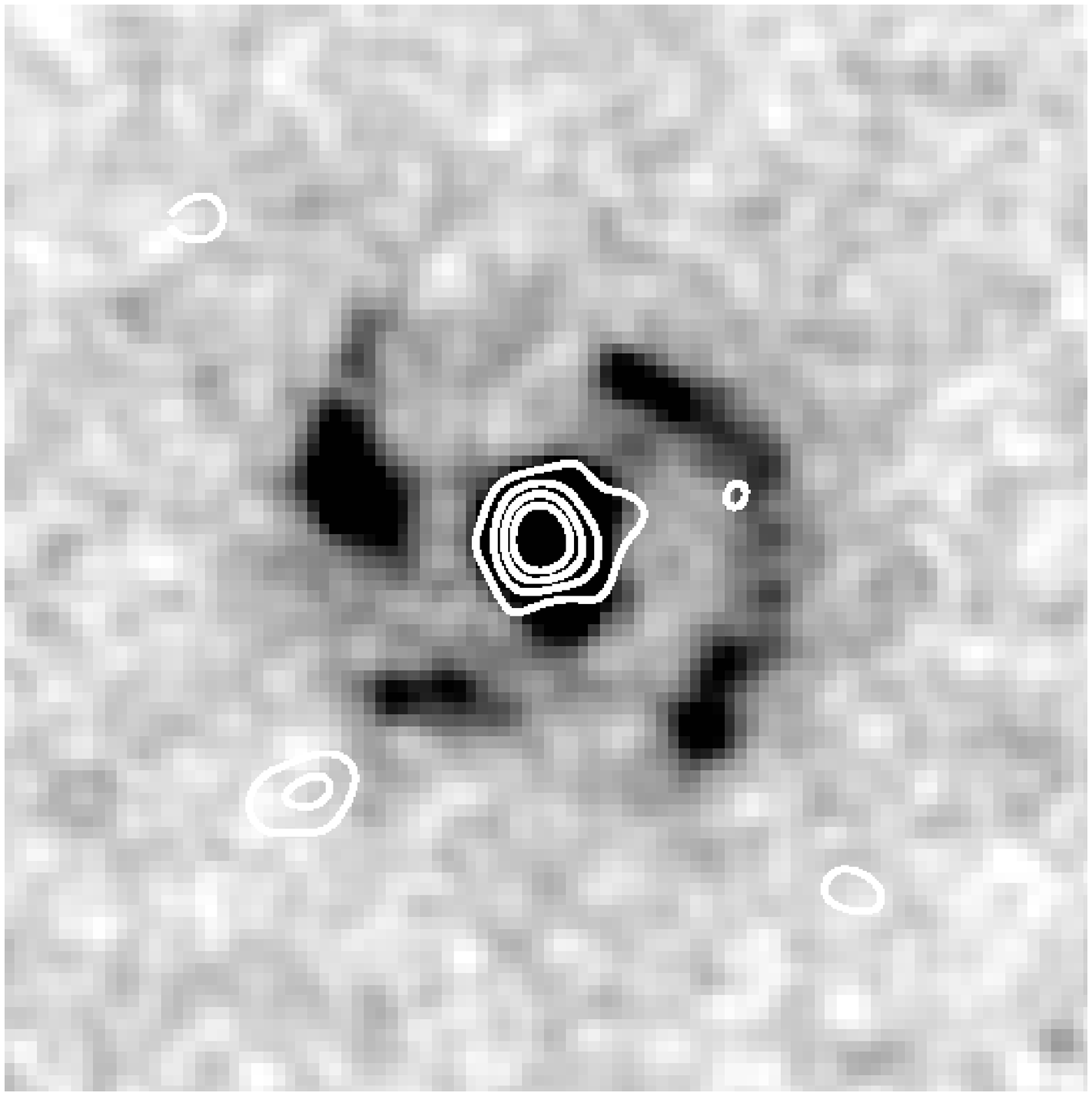,angle=0,width=1.6in,height=1.6in}
\psfig{figure=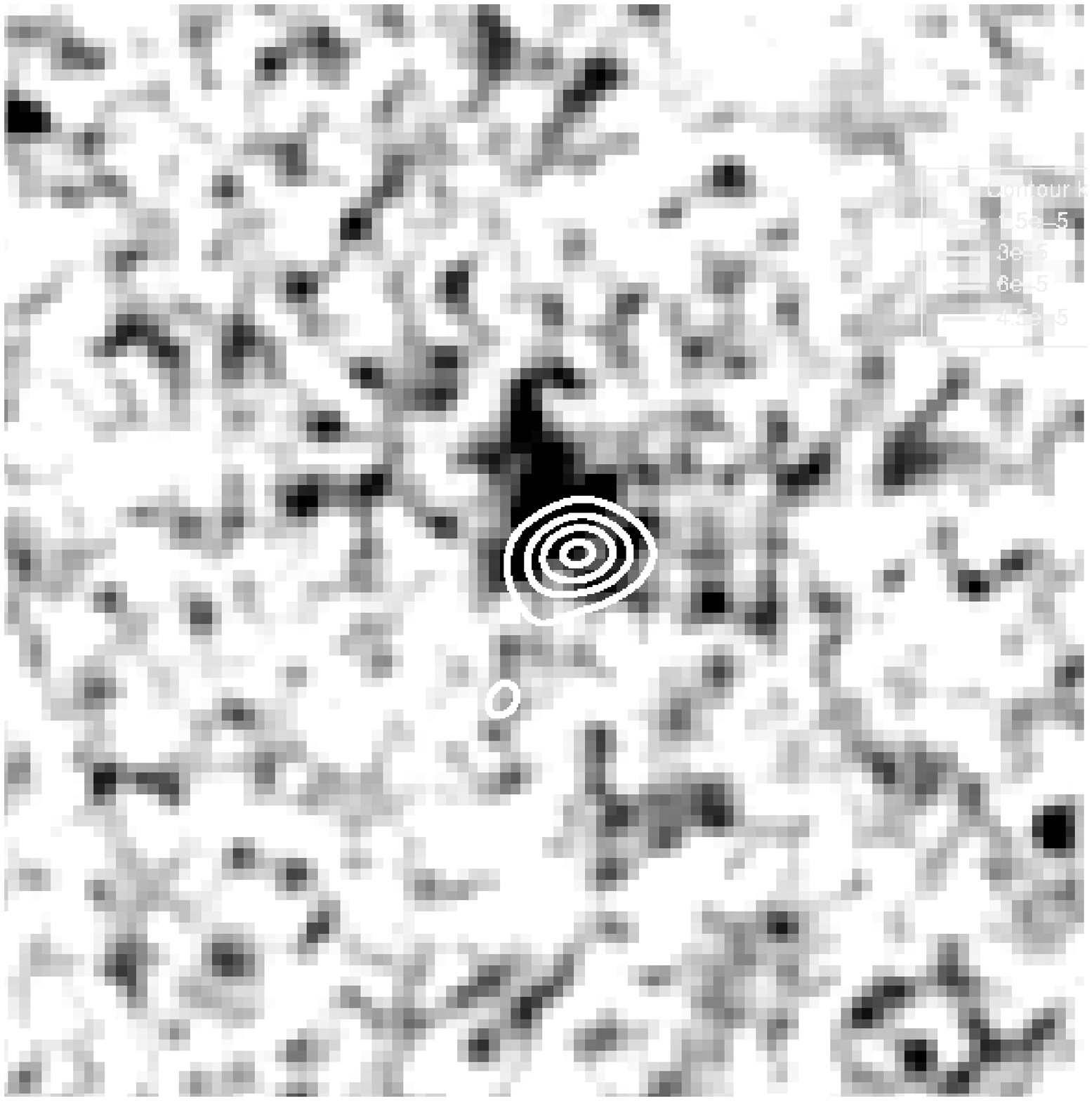,angle=0,width=1.6in,height=1.6in}
}
\caption{\footnotesize
Comparison of MERLIN/VLA radio data (0.2--0.3\arcsec\ synthesized beamsize,
shown as contours) 
for 12 of our SMGs with the {\it HST} STIS {\sc clear50ccd} and 
ACS {$b_{435}$+$i_{775}$} 
imaging (as noted in Table~1).  
The {\it HST} images have been
smoothed with a Gaussian of FWHM 0.2$''$ to match the effective
resolution of the radio map.  The first two rows show
sources where the radio is clearly extended along a portion of the
rest-UV imaging, whereas the final four panels
on the bottom row reveals point sources
in the radio, likely associated with a single UV component. 
Radio contour levels begin at $3\sigma$ and increase in linear steps scaled
to the peak of the radio emission.
Each panel is 3\arcsec\ on a side and they are ordered from top-left as
in the same sense as Table~1.  
}
\label{fig2}
\end{figure*}

\section{Sample and Observations}

Our primary observational dataset is the combined
MERLIN\footnote{MERLIN, a UK National Facility operated by the
University of Manchester at Jodrell Bank Observatory on behalf of
PPARC} and VLA radio map of a $10'\times 10'$ region centered on the
Hubble Deep Field (HDF) region, which has sufficient sensitivity and
resolution to attempt a radio morphological analysis of the submm
galaxies in this area.

The SMG sample in this region comes from the parent catalog of
submm-detected, optically-faint $\mu$Jy radio galaxies used in the
spectroscopic survey of Chapman et al.\ (2003a, 2004).  We identify 14
submm-detected galaxies lying within a $7.5'$-diameter field centered
on 12~36~48.0 +62~15~40 (J2000) which have 850$\mu$m fluxes brighter
than 4\,mJy and are detected at 1.4\,GHz in the VLA A-array
observations of this region with a flux of $>40\mu$Jy.  This radio flux
limit should ensure a useful constraint on the source morphology from
the MERLIN observations.

\subsection{MERLIN Observations}

The deep MERLIN observation of the HDF region (Muxlow et al.\ 1999;
Muxlow et al.\ 2004) comprise $\sim 430$\,hrs integration at 1.4\,GHz
of a $10'\times 10'$ field centered on the HDF and including the Hubble
Flanking Fields (HFF).  These data were acquired in February 1996 and
April 1997.  The MERLIN data were supplemented with 42\,hrs of 1.4-GHz
VLA\footnote{The National Radio Astronomy Observatory is a facility of
the National Science Foundation operated under cooperative agreement by
Associated Universities, Inc.}  A-array observations (Richards 2000),
and combined and deconvolved in the sky-plane due to computational
limitations.  We use a map with a restored 0.3-$''$ beam which has an
rms noise level of 3.3\,$\mu$Jy\,beam$^{-1}$.  To register the radio
and optical data, radio sources associated with compact galaxies have
been used to align the radio map with panoramic ground-based imaging
(see below).  For the HDF, this matching involves 128 $I<24$ optical
sources with radio counterparts and yields an rms of 0.3$''$ (Capak et
al.\ 2004).

\subsection{{\it HST} imaging}

We next search the {\it HST}\footnote{The optical data is based around
observations with the NASA/ESA {\it Hubble Space Telescope}, obtained
at the Space Telescope Science Institute, which is operated by the
Association of Universities for Research in Astronomy, Inc., under NASA
contract NAS5-26555.}  database for deep imaging observations of SMGs
that lie within MERLIN field.  As our primary goal is a comparison of
the coarse morphologies of the sources in the radio and optical
wavebands, the choice of camera used for the {\it HST} observations is
less critical than it would be for a detailed morphological analysis of
SMGs (Chapman et al.\ 2003b). Hence, we search for any observations of
SMGs within the HDF MERLIN field using the STIS and ACS cameras.

Due to the intensive study of this field we identify {\it HST} imaging
of 13 SMGs from our sample which lie in the MERLIN field and list these
in Table~1.  Four of these galaxies come from the {\it HST} targeted
survey of 13 SMGs observed with STIS by Chapman et al.\ (2003b), see
Table~1.  In addition to these, a further 9 SMGs serendipitously fall
in the HDF/HFF region which is covered by ACS imaging from the GOODS
project (Giavalisco et al.\ 2004).

However, in one of these 13 sources (SMM\,J123651.76 +621221.3), the
SMG is not uniquely identified, and may lie behind a foreground
elliptical galaxy (Dunlop et al.\ 2004), and an accurate comparison of
the morphology using the MERLIN/VLA radio and {\it HST} in the optical
is impossible.  We exclude this object from further comparison.

The Cycle 10 STIS imaging of galaxies in our sample uses the open
filter, {\sc clear50ccd} (central wavelength 5733A), and have
durations of two or three orbits (5.0--7.5\,ks).  The reduction and
analysis of these images is described in Chapman et al.\ (2003b).  The
resolution of these images is 0.06$''$ FWHM and they have a typical
point-source sensitivity limit of 27.4~mag.\ (AB).

The ACS observations of the SMGs lying within the GOODS-N field were
obtained from the STScI v1.0 release (August, 2003), in the $i_{775}$
and $b_{435}$ bands.  The reduction of these data is described by
Giavalisco et al.\ (2004) and the typical point-source sensitivity is
28~mag.\ (AB) and resolution of 0.07$''$.

We register the {\it HST} images to the radio coordinate frame by
aligning them with the deep Suprimecam images from Capak et al.\ (2004)
which are tied to the radio frame. To achieve this we first smooth the
{\it HST} images to the ground-based seeing, then match all
$>5$-$\sigma$ sources (except the SMG), and transform the coordinate
grids using the {\sc iraf} task, {\sc geotran}.

The morphological characteristics of this sample have been classified
by eye from the {\it HST} imaging in Fig.~1 and are presented in
Table~1. These rough morphological classes are subject to uncertain
structured dust extinction (Smail et al.\ 1999), and may not represent
the true physical morphology of the system.

\subsection{Archival observations}

Optical photometry for our SMG sample in the $B$ and $R$-bands
(Table~1) was measured from the Subaru Suprimecam imaging published by
Capak et al.\ (2004).  We use a 3\arcsec\ diameter aperture centered on
the radio source and the limiting magnitudes are $B<26.9$ and $R<26.6$
(5$\sigma$).
 
11 of these 12 SMGs in our sample have secure spectroscopic redshifts
from the Keck survey of Chapman et al.\ (2004 -- Table~1).  These
enable us to measure physical sizes and luminosities for these galaxies
and also calculate K-corrections between galaxies in the sample to
directly compare their restframe properties.  $UgRIK$ photometry exists
for the remaining source (SMM\,J123646.1+621449), allowing us to
estimate a photometric redshift.  Using the {\sc hyper-z} software
(Bolzonella et al.\ 2000), we derive a photometric redshift of
$z=1.7\pm 0.2$.  The median redshift of the sample is $<\! z\!>=2.2\pm
0.2$, representative of that measured for a larger spectroscopic
samples of submm galaxies Chapman et al.\ (2003a, 2004).

\section{Analysis and Results}

Using the registered radio and {\it HST} optical imaging we show in
Fig.~1 the 1.4-GHz MERLIN/VLA contours overlayed on the {\it HST}
images for the 12 SMGs in the joint sample.  We note that at the median
redshift of our sample, the {\it HST} imaging corresponds to restframe
wavelengths of 1700--1800A.  The median diameter of the radio
emission from the 12 galaxies is $0.83\pm 0.14''$ or $7.0\pm 1.1$\,kpc
(measured above the 3-$\sigma$ contour, Fig.~1), showing that the
typical source in our sample is well-resolved at the resolution of our
MERLIN/VLA map.  The radio morphologies for these submm galaxies split
into two broad classes: those dominated by an unresolved component,
often centered on a sub-component of the {\it HST}-optical emission
(4/12 or 33\%), and extended structures on scales $\sim$0.5--1\arcsec
(8/12 or 67\%).  Remarkably, the extended radio morphology in these
latter sources often appears to trace the same UV-bright, large-scale
structures seen in the {\it HST} optical images (Fig.~1).

With sensitive, high-resolution imagery in the radio and optical, we
can compare the radio emission (as a tracer of the dust emission) to
restframe-UV emission on kpc-scales within the systems.  In Fig.~2 we
show the radio and restframe UV surface brightness profiles along the
major axes of three representative resolved sources.  In some cases the
UV emission shows broad similarities to the radio emission. However,
Fig.~2 demonstrates that the UV/radio flux ratios can vary
significantly over the extent of the galaxy, and the SFRs derived from
the respective wavelengths will differ accordingly.  For example, in
SMM\,J123621.3+621708 the extended restframe UV emission does not trace
the more compact radio emission, with the peak of the radio emission
actually coinciding with a clear deficit in the UV emission.  Such
anti-correlations in the radio and UV are similar to those seen in many
local ULIRGs (Chamandaris et al.\ 2002).

To test whether there is any evidence for a correlation between
internal reddening and the distribution of obscured star formation
traced by the radio emission, we construct $(b_{435}-i_{775})$ color
maps of the 12 SMGs using the ACS imaging of the GOODS-N region.  Only
SMM\,J123712.0+621325 and SMM\,J123707.2+621408 show redder internal
color structure which corresponds to the radio morphology. In all other
cases, the radio emission does not correspond to any regions with
unusually red colors in the {\it HST} $(b_{435}-i_{775})$ maps.
SMM\,J123707.2+621408 and SMM\,J123712.0+621325 actually show bluer
colors in the vicinity of the peaks in the radio emission.  This
suggests that the UV emission may not always probe the true site of
far-IR emission; and indeed the UV-inferred bolometric luminosities in
SMGs on average underpredict the true luminosities by factors $\gg$10
(Chapman et al.\ 2004).

%
%
\begin{inlinefigure}\vspace{6pt}
\psfig{figure=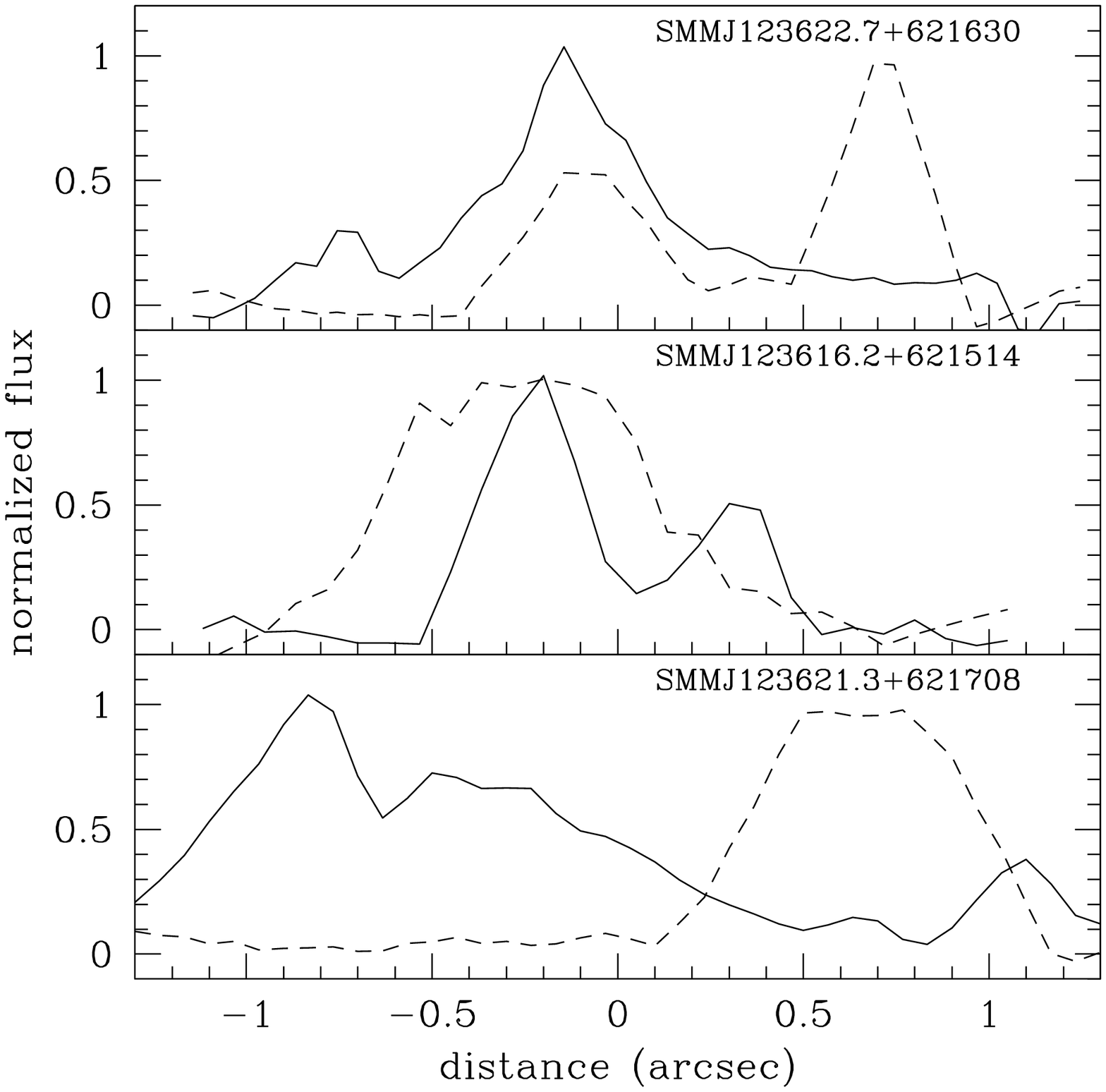,angle=0,width=3.5in}
\vspace{6pt}
\figurenum{2}
\caption{\small
The surface brightness profiles in the radio and optical wavebands
along the major axes of the radio emission in three of the resolved
SMGs.  The dashed line shows the radio emission from the MERLIN/VLA map
and the solid line shows the equivalent distribution of the restframe
UV emission from the {\it HST} imaging. In both cases the surface
brightness is averaged over a 0.3\arcsec\ window orthogonal to the
major axis.  The top axis gives the angular size at the mean redshift
of these three galaxies, the variation in their redshifts produces less
than a 5\% change in angular scale between sources.
}
\label{figslice}
\addtolength{\baselineskip}{10pt}
\end{inlinefigure}

A more quantitative test of the degree of obscuration within these
galaxies is gleaned through comparing the variation in the total
radio/UV flux ratio measurements of each galaxy with that derived
locally for the regions of intense far-IR emission pinpointed by the
MERLIN/VLA morphology.  The monochromatic rest-frame 2000A\
luminosities (L$_{\rm UV}$) of our SMGs are estimated from linear
interpolation between the $B$- and $R$-band magnitudes (Table~1).

We estimate L$_{\rm FIR}$ using the measured radio flux (Table~1),
K-corrected using synchrotron slope of $\alpha=-0.75$ ($S_\nu \propto
\nu^\alpha$) to restframe 1.4\,GHz at the observed redshift, and then
transformed to the far-IR using the local far-IR--radio correlation
from Helou et al.\ (1985).  We take this route to estimate the far-IR
emission on arcsecond-scales as our submm data lack both the spatial
resolution and full spectral information needed to estimate L$_{\rm
FIR}$ more directly.  We note that Garrett (2002) and Kovacs et al.\
(2004) have shown that the far-IR--radion relation does not seem to be
change substantially out to $z\sim1$--2.5, and so the approach we have
adopted should be reliable.  In this manner we predict a median total,
far-IR luminosity of $(3.5\pm 1.0)\times 10^{12}$L$_\odot$ for the 12
SMGs, equivalent to star formation rates of $\sim
1700$M$_\odot$\,yr$^{-1}$ for stars more massive than 0.1\,M$_\odot$
based on a Salpeter IMF.

To investigate the variation of far-IR/UV ratios within the SMGs we
first calculate the ratios for the total emission from the galaxies.
We then measure the same ratio in a region encompassing the peak of the
radio emission using a fixed circular aperature enclosing all radio
emission above the 3$\sigma$ contour shown in Fig.~1.  We compare the
total and locally derived L$_{\rm FIR}$/L$_{\rm UV}$ ratios for the 12
sources in Fig.~3.  We see that when the regions pinpointed as the
sites of strong activity by the MERLIN/VLA radio morphology are
considered, the obscuration levels increase by factors of 2--8 over the
ratios derived for the whole galaxy.  Even measured over the entire
extent of the systems, the obscurations we derive are considerably
higher than those seen in high redshift, restframe UV-selected
populations which are typically L$_{\rm FIR}$/L$_{\rm UV}\sim $0.1--100
(e.g.\ the Lyman-break galaxies, Adelberger \& Steidel 2000; or the
BX/BM galaxies, Steidel et al.\ 2004).

%
%
\begin{inlinefigure}\vspace{6pt}
\psfig{figure=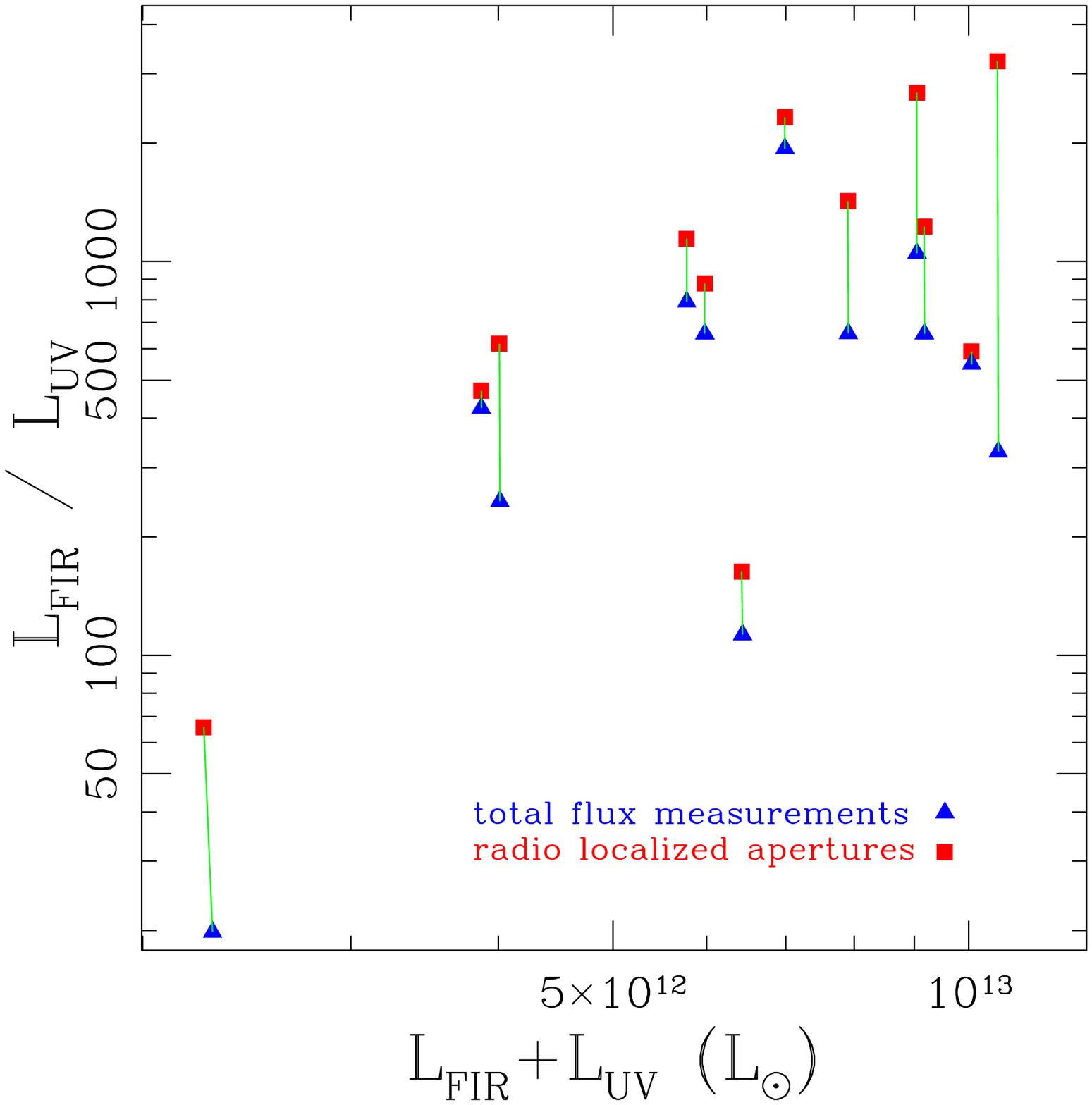,angle=0,width=3.5in}
\vspace{6pt}
\figurenum{3}
\caption{\small
The ratio of far-IR luminosity to restframe UV luminosity (2000A\
monochromatic luminosity) plotted against total luminosity (far-IR +
UV) for the 12 SMGs which are detected in the MERLIN+VLA map. We show
data points corresponding to the ratio derived from total flux
measurements of each source using MERLIN+VLA for radio and {\it HST}
for the restframe UV, and based on small aperture measurements in the
UV and radio based on the positions of the peak of the radio emission
in the MERLIN+VLA map (Muxlow et al.\ 2004).
}
\label{figobs}
\addtolength{\baselineskip}{10pt}
\end{inlinefigure}

\subsection{Searching for very extended radio emission}

Having discovered that the radio emission in over half the SMGs in our
sample is extended on 1\arcsec\ scales we now wish to test whether this
emission extends to even larger scales, $\gs 5$\,\arcsec\ or $\sim
50$\,kpc.  We can constrain the fraction of radio flux arising on $\gs
1.5$\,\arcsec\ scales by a simple comparison of the 1.4-GHz fluxes of
SMGs in maps from the VLA in its B-array (5\arcsec\ synthesized beam)
and A-array (1.5\arcsec\ beam) configurations.  The higher resolution
A-array observations may resolve out a portion of the total emitted
flux density of any source which has significant extended structure on
scales larger than the 1.5\arcsec\ VLA A-array beam.  As no B-array
coverage is available for the HDF-N we have assembled VLA B-array and
A-array fluxes for a sample of 50 $\mu$Jy radio sources, including 5
submm galaxies, from the our SSA22 survey field (see Chapman et al.\
2004).

We calculate the median and bootstrap errors on the flux ratios for the
radio sources from the data taken at the two resolutions.  We start by
confirming that sources show either similar or higher fluxes in the
B-array map compared to the A-array data, as expected considering the
median radio angular sizes of $\mu$Jy sources (Windhorst et al.\ 1993)
We find median ratios of $<\! S_B/S_A\! >=1.45\pm0.27$ and that for
30\% of the sample agree within 2-$\sigma$, suggesting there is no
calibration offset between the fluxes from the two maps.  For the five
submm-detected sources we obtain a median flux ratio of $<\! S_B/S_A\!
>=1.22\pm0.12$, compared to $<\! S_B/S_A\! >=1.22\pm0.11$ for an
$R$-band matched sample of 8 radio sources undetected or unobserved in
the submm waveband.  This suggests that the submm-luminous section of
the $\mu$Jy radio populations is no more extended than the general
population, after we have removed low redshift galaxies (with bright
apparent magnitudes). Moreover, we see that there is only weak evidence
that the radio emission in the SMGs is extended on $\gg 10$\,kpc
scales.

\section{Discussion \& Conclusions}

The far-IR emission from local ULIRGs generally arises from the nuclear
regions of the systems, while the UV light is more extended, although
it contributes a negligible fraction of the bolometric emission
(Goldader et al.\ 2002; Surace \& Sanders 2000).  We have been able to
trace the distribution of these two components in similarly luminous
galaxies at high redshifts through the combination of sub-arcsecond
imagery in the radio and restframe UV from our MERLIN/VLA and {\it HST}
observations.  These maps allowed us to investigate the relative
distribution of obscured and unobscured star formation on kpc-scales
within luminous submm galaxies at $z\sim 2.2$, when the Universe was
only a fifth of its current age.
 
As expected the UV morphologies of this sample are similar to the
(overlapping) sample of submm galaxies imaged with {\it HST}/STIS and
analysed by Chapman et al.\ (2003b), as well as the ACS imaging in
Smail et al.\ (2004).  The sample exhibits irregular and frequently
highly complex morphologies in their restframe $\sim 2000$A\ emission
compared to optically-selected galaxies at similar redshifts, and have
scale lengths far in excess of comparably luminous local galaxies.

Turning to the radio morphologies, we find that in $\sim70$\% (8/12)
the MERLIN/VLA radio exhibits resolved radio emission on $\sim
$0.5--$1''$ scales ($\sim 10$\,kpc) which mirrors the general form of
the restframe UV morphology seen by {\it HST}.  We interpret this as
strong support for the radio emission tracing spatially-extended,
massive star formation within these galaxies.  This situation is very
unlike local ULIRGs, where the high surface brightness far-IR/radio
emission is restricted to a compact nuclear region with an extent of
less than $\sim 1$\,kpc (Charmandaris et al.\ 2002).  A more detailed
analysis of the distribution of UV and radio emission within these
galaxies (Fig.~2 \& 3) shows that the correspondence is rarely
one-to-one, with variations of the UV/radio flux ratio of factors of a
few on kpc-scales within galaxies.  It is likely that comparisons using
even higher resolution data would show even stronger variations, as are
seen in local ULIRGs (e.g.\ Bushouse et al.\ 2002), but which are
diluted at the current resolution.  Similarly, there is only weak
evidence for correlations between the restframe UV colors and the
positions of the radio emitting regions within these galaxies.  Here,
longer wavelength near-IR observations with {\it HST}/NICMOS, or from
the ground, may reveal variations in the UV spectral slope that would
show a better correlation of reddening with radio intensity (Smail et
al.\ 2004).

In the remaining $\sim$30\% (4/12) of SMGs, the radio emission is much
more compact and is essentially unresolved -- suggesting it arises in a
region with a scale size of order $\sim 1$\,kpc or less (Fig.~1).  In
two of these cases, the compact radio emission is centered on a bright
UV source (in one case clearly the nucleus of a face-on spiral galaxy,
which is a strong X-ray source and is also the only one of the sources
in our sample which shows AGN signatures in its UV spectrum), while in
the other two systems the compact radio component is spatially offset
by several kpc from the UV source.  These configurations reflect either
compact, nuclear starbursts and/or a dominant contribution from an AGN
to the radio emission.  In half of these cases the AGN/nuclear
starburst is also strongly obscured at restframe wavelengths of $\sim
2000$A.

Stevens et al.\ (2003, 2004) have recently presented evidence for submm
emission resolved on $\sim100$\,kpc scales (including apparent
filaments) in the rare and extreme environments around powerful radio
galaxies and absorbed QSOs at $z\sim $2--4.  Our results demonstrate
that in many cases ($\sim $70\%) the far-IR emission (as seen in our
MERLIN/VLA radio maps) of the general SMG population is extended on
scales $\sim10$\,kpc.  However, the interferometric measurements are
not suited to measuring larger scale, diffuse emission.  Our VLA
B-array versus A-array comparison (\S3.2), however, suggests that the
typical field SMG does not have submm emission (as traced by the radio)
extending on scales much larger than $\geq$1\arcsec ($\geq 10$\,kpc).

Our high resolution radio and optical imaging allows us to address the
relative obscuration of the galaxies.  Adelberger \& Steidel (2000)
have suggested that high-luminosity galaxies at high redshift have much
stonger obscuration than lower-luminosity galaxies, as measured by the
ratio of far-IR to restframe-UV luminosity. This is certainly true on
large-scales in the submm galaxies.  On smaller scales within the SMGs,
we have seen that the obscuration is roughly $\sim2 \times$ higher over
the region of intense radio (and by implication far-IR) emission,
compared to the average over the whole galaxy.  This suggests that
there is highly structured reddening within the submm galaxies, such
anisotropic obscuration would be a natural consequence of channels
being blown through the dust around the star-formation regions by
vigorous winds.

It is worth considering that few star-burst galaxies with L$_{\rm
FIR}\sim 4\times 10^{12}$\,L$_\odot$ exist locally; most galaxies in
our neighborhood with these luminosities have strong and obvious AGN
components.  However, at $z\sim2.2$, the median redshift of the
radio-selected SMGs, the most active galaxies were evidently forming
stars at rates of $\sim$1700\,M$_\odot$\,yr$^{-1}$ in regions extending
over $\sim$40\,kpc$^2$.  The large physical extent of this activity
contrasts markedly with the compact, nuclear starbursts typical of
local redshift ULIRGs.  This suggests that the some of the
observational properties of the star formation activity in these
galaxies (e.g.\ mix of dust temperatures, ease of superwind generation,
etc.)  may differ markedly from that seen in local ``analogs''.
However, we also note that the star formation surface density inferred
from our radio observations is
$\sim$45\,M$_\odot$\,yr$^{-1}$\,kpc$^{-2}$, comparable to the
upper-limit estimated for such activity in local starburst galaxies by
Meurer et al.\ (1997).  This argues that the small-scale physical
mechanisms which limit the star formation process within these galaxies
are similar to those operating in the most vigorous systems locally.

While evidence for massive amounts of molecular gas in submm galaxies
has now been established (Frayer et al.\ 1998; Neri et al.\ 2003; Greve
et al.\ in preparation), and X-ray luminosities are consistent with a
dominant role for star formation in the energetics of SMGs (Alexander
et al.\ 2004), our discovery of spatially extended radio morphologies
is perhaps the strongest piece of evidence that star formation
dominates the bolometric output of the majority of the submm galaxy
population.

\medskip

In summary, we have compared the restframe UV and radio morphologies on
sub-arcsecond scales of a small sample of highly luminous, dusty
galaxies for which precise redshifts are available.  This analysis
shows that the radio emission, which we adopt as a proxy for the far-IR
emission, in these galaxies is resolved in the majority of galaxies --
implying that dust heating (and by implication, massive star formation)
is occuring on $\sim 10$\,kpc scales within these systems.  Currently,
this represents our only constraint on the likely submm morphology of
these galaxies, and one which will not be further testable until ALMA
comes on-line.  The overall structure of the radio emission matches
that seen in the restframe UV -- although there are strong variations
in the relative emission on kpc-scales -- which we interpret as
resulting from highly structured dust obscuration within the galaxies.
This structured obscuration may reflect from anisotropic dispersal of
the dust as superwinds driven by the star formation activity blows
channels through the intergalactic medium.  Such channels would provide
the opportunity for Ly$\alpha$ photons to escape from these otherwise
highly-obscured systems, explaining the unexpected strength of this
line in their spectra (Chapman et al.\ 2003a, 2004).

\begin{deluxetable}{lccccccl}
\renewcommand\baselinestretch{1.0}
\tablewidth{0pt}
\parskip=0.2cm
\tablenum{1}
\tablecaption{Properties of the submm galaxies}
\small
\tablehead{
\colhead{Source} & {$z$} & {$B$} & {$R$} & {S$_{850}$} & {S$_{1.4}$} & 
\multispan2{~Morphology$^f$~~~~} \\
\colhead{} & {} & {} & {} & {(mJy)} & {($\mu$Jy)} & {Radio} & {Optical}
}
\startdata
SMM\,J123606.9+621021 & 2.509 & 25.6 & 25.2 & 11.6$\pm$3.5 & 74.4$\pm$4.1    & E & Disturbed, merger \\ %
SMM\,J123616.2+621514$^e$ & 2.578 & 26.8 & 25.7 & 5.8$\pm$1.1 & 53.9$\pm$8.4 & E & 3 components \\ 
SMM\,J123622.7+621630$^e$ & 2.466 & 25.6 & 25.4 & 7.7$\pm$1.3 & 70.9$\pm$8.7 & E
& Merging disks \\ 
SMM\,J123629.1+621046$^a$ & 1.013 & 26.1 & 24.6 & 5.0$\pm$1.3 & 81.4$\pm$8.7 & E
& Disturbed \\ 
SMM\,J123655.8+621200$^d$ & 2.743 & 25.4 & 25.3 & 8.0$\pm$1.8 & 21.0$\pm$6.2    & E & Disturbed, merger \\  
SMM\,J123707.2+621408 & 2.484 & 26.9 & 26.0 & 6.3$\pm$1.3 & 45.3$\pm$7.9    & E
& Blue \& red pair  \\  %
SMM\,J123712.0+621325 & 1.992 & 26.0 & 25.8 & 4.1$\pm$1.3 & 53.9$\pm$8.1    & E
& Disturbed with dusty component  \\  %
SMM\,J123712.1+621212$^b$ & 2.914 & 27.0 & 25.5 & 8.0$\pm$1.8 & 21.0$\pm$4.0  & E
& Disturbed, double source \\  
SMM\,J123618.3+621551$^e$ & 1.865 & 26.0 & 25.9 & 7.3$\pm$1.1 & 150.5$\pm$11.2  & C & Small group \\ %
SMM\,J123621.3+621708$^{c,e}$ & 1.988 & 25.1 & 24.9 & 7.8$\pm$1.9 & 148.1$\pm$11.2  & C  & Linear \\  
SMM\,J123635.6+621424 & 2.005 & 24.2 & 24.2 & 5.5$\pm$1.4 & 87.8$\pm$8.8  & C  & Disturbed, face-on spiral \\ %
SMM\,J123646.1+621449 & 1.7 & 25.8 & 25.7 & 10.3$\pm$2.2 & 124.3$\pm$7.9  & C  & Disturbed (photo-z)\\ %
\enddata
\label{tab1}

{\footnotesize
\begin{tabular}{l}
$^a$ The radio source extends over 3.3\arcsec\,
 predominantly to the west of the galaxy. Some very low surface brightness
 emission is\\ missed in the
MERLIN image which just shows three components.
 There is extended emission between these sources.\\
$^b$  Two resolved radio sources trace the two faint UV-detected
sources within a 1\arcsec\ region.\\
$^c$  Two radio sources are found within the SCUBA error
circle, the brighter one listed in the table lacks an optical ID,\\ and the
fainter one associated with a linear galaxy 2\arcsec\ to the east
(see Fig.~1).\\
$^d$ This source was identified in Hughes et al.\ (1998) as HDF\,850.2\\
$^e$ HST imaging from STIS, otherwise with ACS.\\
$^f$ Radio morphologies: extended (E) or compact (C).\\
\end{tabular}}
\end{deluxetable}

\acknowledgements
We would like to thank the anonymous referee for helpful comments which
improved the clarity of the manuscript.  We also thank our collaborator,
Andrew Blain, for his work on the Keck SMG redshift survey.
Support for proposal \#9174 (SCC, RW) was provided by NASA through a
grant from the Space Telescope Science Institute, which is operated by
the Association of Universities for Research in Astronomy, Inc., under
NASA contract NAS5-26555.  IRS acknowledges support from the Royal
Society and Leverhulme Trust.

\end{document}